
\input harvmac
\def\CS{{\cal S}}

\def\inbar{\,\vrule height1.5ex width.4pt depth0pt}
\def\IB{\relax{\rm I\kern-.18em B}}
\def\IC{\relax\hbox{$\inbar\kern-.3em{\rm C}$}}
\def\ID{\relax{\rm I\kern-.18em D}}
\def\IE{\relax{\rm I\kern-.18em E}}
\def\IF{\relax{\rm I\kern-.18em F}}
\def\IG{\relax\hbox{$\inbar\kern-.3em{\rm G}$}}
\def\IH{\relax{\rm I\kern-.18em H}}
\def\II{\relax{\rm I\kern-.18em I}}
\def\IK{\relax{\rm I\kern-.18em K}}
\def\IL{\relax{\rm I\kern-.18em L}}
\def\IM{\relax{\rm I\kern-.18em M}}
\def\IN{\relax{\rm I\kern-.18em N}}
\def\IO{\relax\hbox{$\inbar\kern-.3em{\rm O}$}}
\def\IP{\relax{\rm I\kern-.18em P}}
\def\IQ{\relax\hbox{$\inbar\kern-.3em{\rm Q}$}}
\def\IR{\relax{\rm I\kern-.18em R}}
\font\cmss=cmss10 \font\cmsss=cmss10 at 7pt
\def\IZ{\relax\ifmmode\lrefhchoice
{\hbox{\cmss Z\kern-.4em Z}}{\hbox{\cmss Z\kern-.4em Z}}
{\lower.9pt\hbox{\cmsss Z\kern-.4em Z}}
{\lower1.2pt\hbox{\cmsss Z\kern-.4em Z}}\else{\cmss Z\kern-.4em Z}\fi}
\def\IGa{\relax\hbox{${\rm I}\kern-.18em\Gamma$}}
\def\IPi{\relax\hbox{${\rm I}\kern-.18em\Pi$}}
\def\ITh{\relax\hbox{$\inbar\kern-.3em\Theta$}}
\def\IOm{\relax\hbox{$\inbar\kern-3.00pt\Omega$}}

\def\np{Nucl. Phys. }
\def\pl{Phys. Lett. }

\def \CA{{\cal A}}

\def \CA{{\cal A}}

\def \ab{\bar{a}}

\def \l{\ell}

\def \sinh{{\rm sinh}}
\def \cosh{{\rm cosh}}

\newdimen\xraise\newcount\nraise
\def\xpoint{\hbox{\vrule height .45pt width .45pt}}
\def\udiag#1{\vcenter{\hbox{\hskip.05pt\nraise=0\xraise=0pt
\loop\ifnum\nraise<#1\hskip-.05pt\raise\xraise\xpoint
\advance\nraise by 1\advance\xraise by .4pt\repeat}}}
\def\ddiag#1{\vcenter{\hbox{\hskip.05pt\nraise=0\xraise=0pt
\loop\ifnum\nraise<#1\hskip-.05pt\raise\xraise\xpoint
\advance\nraise by 1\advance\xraise by -.4pt\repeat}}}
\def\bdiamond#1#2#3#4{\raise1pt\hbox{$\scriptstyle#2$}
\,\vcenter{\vbox{\baselineskip12pt
\lineskip1pt\lineskiplimit0pt\hbox{\hskip10pt$\scriptstyle#3$}
\hbox{$\udiag{30}\ddiag{30}$}\vskip-1pt\hbox{$\ddiag{30}\udiag{30}$}
\hbox{\hskip10pt$\scriptstyle#1$}}}\,\raise1pt\hbox{$\scriptstyle#4$}}
\def\CD {{\cal D}}

\def\p {\partial}
\def\CS {{\cal S}}

\def\inbar{\,\vrule height1.5ex width.4pt depth0pt}
\def\IB{\relax{\rm I\kern-.18em B}}
\def\IC{\relax\hbox{$\inbar\kern-.3em{\rm C}$}}
\def\IP{\relax{\rm I\kern-.18em P}}
\def\IR{\relax{\rm I\kern-.18em R}}
\def\inbar{\,\vrule height1.5ex width.4pt depth0pt}
\def\IB{\relax{\rm I\kern-.18em B}}
\def\IC{\relax\hbox{$\inbar\kern-.3em{\rm C}$}}
\def\ID{\relax{\rm I\kern-.18em D}}
\def\IE{\relax{\rm I\kern-.18em E}}
\def\IF{\relax{\rm I\kern-.18em F}}
\def\IG{\relax\hbox{$\inbar\kern-.3em{\rm G}$}}
\def\IH{\relax{\rm I\kern-.18em H}}
\def\II{\relax{\rm I\kern-.18em I}}
\def\IK{\relax{\rm I\kern-.18em K}}
\def\IL{\relax{\rm I\kern-.18em L}}
\def\IM{\relax{\rm I\kern-.18em M}}
\def\IN{\relax{\rm I\kern-.18em N}}
\def\IO{\relax\hbox{$\inbar\kern-.3em{\rm O}$}}
\def\IP{\relax{\rm I\kern-.18em P}}
\def\IQ{\relax\hbox{$\inbar\kern-.3em{\rm Q}$}}
\def\IR{\relax{\rm I\kern-.18em R}}
\font\cmss=cmss10 \font\cmsss=cmss10 at 7pt
\def\IZ{\relax\ifmmode\mathchoice
{\hbox{\cmss Z\kern-.4em Z}}{\hbox{\cmss Z\kern-.4em Z}}
{\lower.9pt\hbox{\cmsss Z\kern-.4em Z}}
{\lower1.2pt\hbox{\cmsss Z\kern-.4em Z}}\else{\cmss Z\kern-.4em Z}\fi}
\def\IGa{\relax\hbox{${\rm I}\kern-.18em\Gamma$}}
\def\IPi{\relax\hbox{${\rm I}\kern-.18em\Pi$}}
\def\ITh{\relax\hbox{$\inbar\kern-.3em\Theta$}}
\def\IOm{\relax\hbox{$\inbar\kern-3.00pt\Omega$}}

\def\l {\ell }

\def\p {\partial}
\def\CS {{\cal S}}

\def\CZ {{\cal Z}}
\def\CA {{\cal A}}

\def\CD {{\cal D}}

\def\pl{Phys. Lett. B }
\def\np{Nucl. Phys.B }


\def\R{\relax{\rm I\kern-.18em R}}
\font\cmss=cmss10 \font\cmsss=cmss10 at 7pt
\def\Z{\relax\ifmmode\mathchoice
{\hbox{\cmss Z\kern-.4em Z}}{\hbox{\cmss Z\kern-.4em Z}}
{\lower.9pt\hbox{\cmsss Z\kern-.4em Z}}
{\lower1.2pt\hbox{\cmsss Z\kern-.4em Z}}\else{\cmss Z\kern-.4em Z}\fi}


\lref\Iml{I.K. Kostov, \pl B 266 (1991) 42}
\lref\Icar{I.Kostov, ``Strings embedded in Dynkin diagrams'',  Lecture
given at the Cargese meeting, Saclay preprint
SPhT/90-133.}
\lref\bkz{E. Br\'ezin, V. Kazakov, and Al. B. Zamolodchikov,
\np 338 (1990) 673}
\lref\Iade{I. Kostov, \np  326 (1989)583.}
\lref\Inonr{I. Kostov, \pl 266 (1991) 317.}
\lref\df{V.Dotsenko and V. Fateev, \np 240 (1984) 312}
\lref\Imult{I. Kostov,
\pl 266 (1991) 42.}
\lref\ajm{J. Ambjorn, J. Jurkiewicz and Yu. Makeenko, \pl 251 (1990)517}
\lref\mat{V. Kazakov, \pl 150 (1985) 282;
F. David, \np 257 (1985) 45;
V. Kazakov, I. Kostov and A.A. Migdal, \pl 157 (1985),295;
J. Ambjorn, B. Durhuus, and J. Fr\"ohlich, \np 257 (1985) 433}
\lref\brkz{E. Br\'ezin and V. Kazakov, \pl 236 (1990) 144.}
\lref\Ion{I. Kostov, {\it Mod. Phys. Lett.} A4 (1989) 217}
\lref\dglsh{M. Douglas and S. Shenker, \np B 335 (1990) 635.}
\lref\grmg{D. Gross and A. Migdal, {\it Phys. Rev. Lett.} 64 (1990) 127.}
\lref\mike{M. Douglas, \pl 238 (1990) 176.}
\lref\pol{A. Polyakov, \pl 103  (1981) 207, 211.}
\lref\kpz{V. Knizhnik, A. Polyakov and A. Zamolodchikov,{\it Mod. Phys. Lett.}
A3 (1988) 819.}
\lref\ddk{F. David,{\it Mod. Phys. Lett.} A3 (1988) 1651; J. Distler and
H. Kawai, \np 321 (1989) 509}
\lref\bk{M. Bershadski, I. Klebanov,\np 360 (1991) 559. }
\lref\mss{G.Moore, N.Seiberg and M. Staudacher, \np 362 (1991) 665.}
\lref\polci{J. Polchinski, \np 346 (1990) 253 }
\lref\stau{M. Staudacher, \np 336 (1990) 349.}
\lref\gm{D. Gross and A. Migdal, \np340 (1990) 333}
\lref\brkz{E. Br\'ezin and V. Kazakov, \pl  236 (1990) 144}
\lref\dglsh{M. Douglas and S. Shenker, \np 335 (1990) 635}
\lref\grmg{D. Gross and A. Migdal, {\it Phys. Rev. Lett.} 64 (1990) 127}
\lref\mike{M. Douglas, \pl 238 (1990) 176}
\lref\pol{A. Polyakov, \pl 103 (1981) 207, 211}
\lref\nati{N. Seiberg,``Notes on Quantum Liouville Theory and Quantum
Gravity,'' Proceedings of the 1990 Yukawa International Seminar, {\it
Prog.Theor. Phys. Supp.} 102 }
\lref\cmnp{A. Cohen, G. Moore, P. Nelson, and J. Polchinski,
\np 247(1986) 143;
A. Cohen, G. Moore, P. Nelson, and J. Polchinski,
``An Invariant String Propagator,'' in {\it Unified
String Theories}, M. Green and D. Gross, eds. World Scientific,
1986}
\lref\moore{G. Moore, \np 368 (1992) 557}
\lref\gr{I.S. Gradshteyn and I.M. Ryzhik, {\it Table of Integrals, Series
and Products}, Academic Press, 1980.}
\lref\ab{M. Abramowitz and I. Stegun, {\it Handbook of Mathematical
Functions}, Dover, 1972.}
\lref\lizu{B. Lian and G. Zuckerman, ``New Selection Rules and Physical
States in 2d Gravity,'' Yale preprint YCTP-P18-90}

\lref\gkl{ D. Gross and I.Klebanov, \np 344 (1990) 475}
\lref\irrationalivan{I.K. Kostov, \pl 266 B (1991) 317}
\lref\Imult{I.K. Kostov, \pl 266B (1991) 42}
\lref\ivanlat{I.K. Kostov,  \np 376 (1992) 539}
\lref\Iade{I.K.Kostov, \np 326 (1989) 583}
\lref\dfkt{P. Di Francesco and D. Kutasov, \pl 261 B (1991) 385;
P. DiFrancesco and D. Kutasov, \np 375 (1992) 119}
\lref\kusei{D. Kutasov and N. Seiberg, \np 358 (1991) 600}
\lref\bk{M. Bershadski  and I. Klebanov, \np 360 (1991) 559}
\lref\coul{ B. Nienhuis, {\it in Phase transitions
and critical Phenomena} , Vol. 11, ed. C.C. Domb
and J.L. Lebowitz (Academic Press, New York, 1987) ch. 1;
 P. Di Francesco, H. Saleur and J.-B. Zuber,
{\it J. Stat. Phys.} 49 (1987) 57 }
\lref\msei{G. Moore and N. Seiberg, {\it Int. Journ. Mod. Phys.}
A7 (1992) 2601}
\lref\kami{V. Kazakov and A. Migdal, \np 311 (1989) 171}
\lref\daje{S. R. Das and A. Jevicki, {\it Mod. Phys. Lett. }A5 (1990) 1639}
\lref\mqma{E. Br\'ezin, V. Kazakov and Al.B. Zamolodchikov,
\np 338 (1990) 673; D.J. Gross and N. Miljkovic, \pl 238 B (1990) 217;
G. Parisi, \pl 238 B (1990) 209; P. Ginsparg and J. Zinn-Justin,
\pl 240 B (1990) 333}
\lref\mqmb{D.J. Gross and I.R. Klebanov, \np 344 (1990) 475;
\np 352 (1991) 671; \np 354 (1991) 459; \np 359 (1991) 3}
\lref\mqmc{For a review containing further references see:
I.R. Klebanov,``String Theory in Two Dimensions'',
1991 Trieste lectures, Princeton preprint PUPT-1271}
\lref\mqmd{Latest developments are described and referenced in:
R. Dijkgraaf, G.Moore and R.Plesser, ``The Partition Function of 2D
String Theory'', Princeton-Yale preprint IASSNS-HEP-92/48,
YCPT-P22-92}
\lref\IMat{I. Kostov and
 M. Staudacher, Multicritical phases
of the $O(n)$ model on a random lattice, Saclay preprint SPhT/92/025,
to appear in \np }.
\lref\ivmm{I.K. Kostov, ``Gauge invariant matrix model for A-D-E closed
strings'', to be submitted to \pl }

\Title{RU-92-21}
{\vbox{\centerline
{Strings in Discrete and Continuous Target Spaces:}
\vskip2pt
\centerline{
A Comparison}}}

\vskip6pt

\centerline{Ivan K. Kostov \footnote{$ ^\ast $}{on leave of absence
from the Institute for Nuclear Research and Nuclear Energy,
Boulevard Tsarigradsko shosse 72, BG-1784 Sofia, Bulgaria}}

\centerline{{\it Service de Physique Th\'eorique
\footnote{$ ^\dagger$}{Laboratoire de la Direction des Sciences
de la Mati\`ere du Comissariat \`a l'Energie Atomique} de Saclay
CE-Saclay, F-91191 Gif-Sur-Yvette, France}}

\bigskip\centerline{Matthias Staudacher}

\centerline{{\it Department of Physics and Astronomy,
Rutgers University, Piscataway, NJ 08855-0849}}

\vskip .3in

\baselineskip10pt{
We find the precise relationship between the loop gas
method and the matrix quantum mechanics approach to two-dimensional
string theory. The two systems are distinguished by different target
spaces ($\Z$ and $\R$, respectively) as far as {\it observables} are concerned.
We argue that target space loop correlators should
coincide in the two models and demonstrate this for a number of examples.
As a consequence some interesting generic observations about the
structure of two-dimensional string theory may be made:
Restricting to a discrete target space leads to {\it factorization}
of amplitudes and thus to very simple sewing rules.
It is also demonstrated that the restriction to the discrete target space
still allows to calculate the correlation functions of
tachyon operators in the unrestricted theory.
}

\vskip 1cm
\leftline{submitted for publication to {\it Physics Letters B}}
\rightline{SPhT/92-092}
\Date{8/92}

\baselineskip=20pt plus 2pt minus 2pt

\newsec{Introduction}

Over the past few years dramatic progress towards formulating a theory
of low-dimensional bosonic strings has been achieved. In fact, for
non-critical strings
living in space-times of dimensions less or equal than one the theory could
be considered solved: One is now able to calculate almost all
physical quantities using lattice models which are exactly solvable
through matrix models or combinatorial techniques. However, possessing the
solution does not necessarily mean one {\it understands} it. Consider
for example the KPZ scaling relation \kpz : It is certainly confirmed by all
exact solutions but in order to give a universal and simple argument
for its validity one should use continuum reasoning \ddk.
Now there clearly exists a host of non-trivial results beyond mere scaling
laws which simply ``come out'' of the exact solutions but are not yet
derivable from a continuum theory. It is in precisely this sense
that the lattice model solutions have been termed ``experimental'' by
some authors.
Thus, the ultimate theory explaining and unifying
all approaches is still missing.

In this situation it is important to carefully analyze {\it all} the
available data. In the present work we aim to contribute to this program
by relating two rather different lattice models designed to discretize
the non-critical string in one dimension. The first model is
formulated as the quantum mechanics of a large $N$ hermitian
matrix \kami. It can be rewritten as an intriguingly simple
system of free fermions which upon bosonization turns into
a rather unusual string field theory with only one interaction
vertex \daje. The model has been exhaustively solved in
references \mqma, \mqmb, \moore, \mqmc, \mqmd
(and references therein). The method of solution
appears however rather removed from more traditional approaches
to string field theory which uses factorization, sewing and
infinitely many interactions. It is precisely these concepts
which appear naturally in the second model we are discussing: The
SOS string \irrationalivan, \Imult , \ivanlat. Here the integrable models where
the target space is an extended\foot{
The techniques apply
equally to the case of the non-extended diagrams ADE corresponding
to $C<1$ systems \Iade . In fact, one of the strenghts of the loop gas
approach is a {\it unified} description of all $C\leq$
noncritical strings. In the present work we restrict the discussion
to $C=1$.}
${\hat A}{\hat D}{\hat E}$
Dynkin diagram are adapted to dynamical lattices.
These extended diagrams correspond to special points in the
space of $C=1$ theories. In particular the $n \rightarrow \infty$
limit of the ${\hat A}_n$ model (i.e. ${\hat A}_n
\rightarrow \Z$) turns into the non-compact $C=1$ system.
It was therefore quite dissatisfying to observe that not only
the methods, but also the results of the SOS approach seemed to be
incompatible with the matrix quantum mechanics; e.g. the
correlator of two macroscopic loops (the propagator) seemed
to be distinct in the two systems. The main result of this
paper is a reconciliation of these differences. It will
be argued that both models indeed describe the same
continuum string theory, albeit in a rather different manner.

The outline of the paper is as follows: In the next section we
give a short
description of the matrix quantum mechanics method as well as the
SOS string models which we also call $\R$ and
$\Z$ strings, respectively.
In section 3 we compare the two systems using both known
results as well as some novel calculations.
Our conclusions and some speculations are presented in section 4.

\newsec{A brief description of the models}

Both models can be considered as discretizations of the Polyakov path
integral
\eqn\polii{\CZ= \int
\CD x \CD g\ e^{-\CA[x,g]}}
\eqn\aaca{\CA[x,g]={1 \over 4\pi }\int d^{2}\xi
 \sqrt{{\rm det} g(\xi)}\ g^{ab}(\xi) \p_{a}x(\xi) \p_{b}x(\xi)
 + \mu \int d^{2}\xi
\sqrt{{\rm det} g(\xi)}}
where the parameter $\mu$  coupled to the   intrinsic area
of the world sheet is the cosmological constant.
%
The measure over the metrics $g_{ab}(\xi)$ is discretized in both cases by
planar graphs but the embedding into the  $x$-space is constructed
differently.

\noindent
$a$)  $\R$-string (Matrix quantum mechanics)

The path integral of the $\R$-string
is given by the sum of all $\varphi^{3}$ planar graphs embedded in the
continuous line $\R$.
Each point $ s$ of the graph $\CS$ has a coordinate $x( s)$ and the
 partition function
is obtained by summing over all possible graphs and integration over the
coordinates of the points. The weight of an embedded
graph $\CS \to \R$ is composed by a topology-dependent factor $N^{\chi}$
where $\chi [\CS]$ is the Euler characteristic of the graph, and a product of
local factors $\Omega_{x( s) , x( s ')}$ associated with the edges
 $< s s'>$ of the graph
\eqn\omomm{\Omega_{x,x'}= e^{- \beta -  |x-x'| }}

The partition function reads
\eqn\pprts{\CZ_{\R} =\sum_{ \CS}N^{\chi[ \CS]}
 \prod_ {s\in \CS } \int_ {-\infty}^{\infty}d x( s)
\prod_{< s s'> \in \CS}
\Omega_{x(s), x( s')} }

The sum over graphs with given topology is convergent for $\beta $
larger than some critical $\beta ^{*}$;
the difference $\beta - \beta ^{*}$ is proportional to the renormalized
cosmological constant $\mu$.

\noindent
$b$) $\Z$-string (SOS model on a random lattice)

The SOS string has as a target space the discretized line $\Z$.
Its path integral is defined as the sum of triangulated surfaces
embedded in $\Z$. An embedded surface
is described by a triangulation (the world sheet) and an
integer valued local field variable (height)
$x(s) \in \Z$ associated with each site $s$ of
the triangulation.  The rules of embedding are such that the  the
heights of the endpoints of each bond $<ss'>$
either coincide or differ by 1.
The weight of an embedded surface $\CS \to \Z$
 depends on its Euler characteristics   $\chi [\CS]$
through the factor $N^{\chi [\CS]}$. Apart of this it is a product
of factors $\Omega _{x(s),x(s')}$ associated with the
bonds $<ss'>$ of the triangulation $\CS$
\eqn\omomg{\Omega_{xx'}=e^{-\beta } \delta _{x,x'} + e^{-\kappa}
[\delta_{x,x'+1}+ \delta_{x,x'-1}]}
The partition function is defined therefore as
\eqn\prtf{\CZ_{\Z}=\sum_{\CS}N^{\chi[\CS]}
\sum_ {\{ x(s) \in \Z \ |s\in \CS \} }
\ \ \prod_{< s s'> \in \CS}
\Omega_{x(s), x( s')} }
The continuum limit is achieved along a critical
 line in the $\beta, \kappa$ space \ivanlat .

\smallskip

\newsec{Comparison}

\subsec{Torus}

The torus diagram is of prime importance since it gives information
about the states of the theory \kusei, \bk. Moreover, it has been
calculated in continuum Liouville theory \bk. The result for
$C=1$ matter compactified on a circle of radius $R$
is
\eqn\tor{\CZ_{{\rm torus}} (R) =- {1 \over 24} (R + {1 \over R}) \log \mu}
This is precisely the result obtained from the $\R$-string with
compact target space \gkl.

  In the $\Z$-string the radius of the circle
can take only integer values $R=h; \  h=1,2,... $ . (We took into account
the scale factor $\pi$ between the two $x$-spaces.)  The corresponding
compactified target space $\Z_{2h}$
is constructed as a closed chain of $2h$ points and is identical
to the  $\hat{A}_{2 h-1}$ Dynkin diagram \foot{Generally it makes sense to
consider only target spaces with even number of points; otherwise winding
modes are kinematically impossible. There is however one exception: the
space $\Z_{1}$ which corresponds  to $R=1/2$. The $\Z_{1}$ string
 is identical to the $O(2)$ model on a random lattice
\Ion }.
The calculation is done in \ivanlat\ and one again finds
 eq.\tor.
{\it We therefore conclude that $\R$ and $\Z$ strings describe the
same continuum string theories in the bulk}. The reader may thus
wonder whether the name $\Z$ string is really appropriate since we
just argued that the discreteness of the target space vanishes in the
continuum limit: The Dynkin diagram turns into a continuous space in
much the same way an Ising model renormalizes onto a continuous
field. We will however see shortly that a remnant of the
discreteness of the target space of the lattice models
 remains
in the continuum once we introduce
boundaries into the manifolds. Before turning to the cylinder
where this effect appears let us first review the simplest case of
a diagram with boundary: The disc.

\subsec{ One loop (Disc)}
The disc amplitude can be formally considered as the mean value of the
operator $w(\l ,x)$ creating on the world sheet a boundary of length $\l$
and position $x$ in the target space.
In the continuum approach the disc amplitude
 should be given
by a product of three separate path integrals over matter, ghost
and Liouville sectors:
\eqn\psif{ \langle w(\l , x)\rangle
=\CZ_{{\rm ghost}} \CZ_{{\rm matter}}(x)  \CZ_{{\rm Liouville}}(\l)}
The reason for this ``factorization'' is that there are no moduli
on the disc. 
%
The path integral for  $\CZ_{{\rm matter}}(x) $ with a Dirichlet boundary
condition for the position field actually does not depend on its boundary
value $x$ since our target space is translationary invariant.
 It is not known yet how to calculate the path-integral for
$\CZ_{{\rm Liouville}}(\l)$.
Indeed it is not clear how to properly treat boundaries in
Liouville theory. Aside from the technical problem of
carrying out the integration one first has to understand
the correct boundary conditions on the Liouville and ghost fields.
In \mss \ qualitative arguments were given that at $C=1$,
and for Dirichlet boundary conditions on the matter field $w(\l)$
should
satisfy the Wheeler-deWitt equation
$[-(\l \partial_\l)^2 + \mu \l^2 +1][\l w_{p}(\l)] = 0$
which  is solved by
\eqn\kzero{\langle w(\l , x ) \rangle = {\sqrt{\mu}
 \over \l} K_{1}(\sqrt{\mu} \l) }
where $K_1$ is a modified Bessel function.
This is indeed what one finds in both the $\R$ \mss,\moore\  and the
$\Z$ string \Icar \ivanlat .

\subsec{Two loops (Cylinder)}

The cylinder amplitude is the tree level propagator of the two-dimensional
string theory. It should be given in the continuum by
the formal path integral
\eqn\prop{\eqalign{
&
\langle w(\l_1 ,x_{1}) w(\l_2 , x_{2}) \rangle =\cr
&
\l_1 \l_2 \int_{0}^{\infty} {d \tau \over \tau } Z_{{\rm matter}}(\tau)
Z_{{\rm ghosts}}(\tau) Z_{{\rm  Liouville}}(\tau,\l_1,\l_2)\cr }}
where $\tau$ is the modular parameter of the cylinder
 playing the r\^ole of a proper time
for the closed string and $\l_1$,$\l_2$ are the
lenghts of the two boundaries. 
%
%
Again this path-integral has not been directly calculated but may
be computed with relative ease in the discrete approach.
 Let us now discuss
the $\R$ and $\Z$ strings separately.

 In the $\R$ case we may place the
boundaries at arbitrary positions $x_1$,$x_2$ in target space.
Then one Fourier-transforms to momentum space
\eqn\four{\delta(q_1 + q_2) \langle w_{q_1}(\l_1) w_{-q_1}(\l_2) \rangle =
\int_{-\infty}^{\infty} dx_1 e^{i q_1 x_1}
\int_{-\infty}^{\infty} dx_2 e^{i q_2 x_2}
\langle w(\l_1,x_1) w(\l_2,x_2) \rangle }
and obtains   \moore, \mss\
\eqn\macror{\eqalign{
&\langle w_q(\l_1) w_{-q}(\l_2) \rangle =\cr
&={\pi q \over \sin \pi q} I_q(\sqrt{\mu} \l_1) K_q(\sqrt{\mu} \l_2)
+ \sum_{r=1}^{\infty} {2 (-1)^r r^2 \over r^2-q^2} I_r(\sqrt{\mu}\l_1)
K_r(\sqrt{\mu}\l_2) } }
where $I$ and $K$ are modified Bessel functions.
Introducing the complete system of $\delta$-function
normalized eigenstates of the kernel \macror\
\eqn\eiggs{ \langle \l |E\rangle ={2 \over \pi}
\sqrt{E \sinh(\pi E)}\  K_{iE}(\sqrt{\mu}\l) , \ \ \ E>0}
one can represent it as an integral
\eqn\ddiaa{\langle w_q(\l_1) w_{-q}(\l_2) \rangle =
\int_{0}^{\infty} dE \ \langle \l_{1}|E\rangle {1 \over E^2 + q^2} \
 {E \over \sinh (\pi E)} \langle E|\l_{2} \rangle}
with $E$ playing the role of the momentum
associated with the Liouville-mode.
The representation as a discrete sum \macror\ is to be interpreted as
a sum over on-shell (microscopic) states of the closed string.

Let us turn to the case of the $\Z$-string.
We argue that in view of \prop\ and the Gaussian nature of
the matter field one should obtain the {\it same} target space correlator
$\langle w(\l_1,x_1) w(\l_2,x_2) \rangle$ as in the $\R$ case.
However, there is one important difference: In the loop gas
formulation it is impossible to choose $x_1$,$x_2$ arbitrary:
{\it The target space distance $|x_2 - x_1|$ does not
renormalize.} Since $x_1$,$x_2$ take on integer values in the
Dynkin diagram they have to remain integers even {\it after} taking the
continuum limit. This ``nonrenormalization'' effect in SOS models
(which is a particuliarity of the statistical model and has
nothing to do with 2D gravity) is e.g. explained in \coul.
It immediately follows that we cannot transform to
momentum space using \four. What remains possible is to transform
to a compact  momentum space  $- 1 <p< 1$ dual to our discrete target space.
The analog of eq.\four\ is then \foot{
We will only discuss the noncompact case $\Z_{2h}, \ h \to \infty$
when  the target  space
 becomes the set of integers $\Z$.
For compact target space ($h$ finite)
 the $p$-momentum space
is compact {\it and} discrete: $p=0,\pm {1 \over h},\pm {2 \over h},
1$.}
\eqn\fouz{\delta^{(2)}(p_1 + p_2) \langle w_{p_1}(\l_1) w_{p_2}(\l_2) \rangle =
\sum_{x_{1} \in \Z } e^{i \pi p_1 x_1}
\sum_{x_{ 2} \in \Z}e^{i \pi p_2 x_2}
\langle w(\l_1,x_1) w(\l_2,x_2) \rangle }
Here $\delta^{(2)}$ is a periodic delta function of period $2$.
In this $p$-momentum space it has been shown shown \ivanlat\ that
\eqn\macroz{\eqalign{
\langle w_p(\l_1) w_{-p}(\l_2) \rangle &=
\int_{0}^{\infty} dE
\langle\l_1 |E \rangle {1 \over \cosh \pi E - \cos \pi p}
\langle E|\l_{2} \rangle \cr
&={1 \over \sin \pi |p|} [
\sum_{n=-\infty}^{\infty} (|p| + 2n)
I_{||p|+2n|}(\sqrt{\mu}\l_1)
K_{|p|+2n}(\sqrt{\mu}\l_2) ]} }
Now let us demonstrate that the two propagators \macror, \macroz\
indeed coincide in $x$-space. To prove this we simply
calculate the inverse Fourier  transforms; first for the $\Z$ case
\eqn\tarz{ {1 \over 2} \int_{-1}^{1} dp {e^{i \pi p x} \over
\cosh \pi E - \cos \pi p} =  {1 \over \sinh \pi E} e^{-E \pi |x|}
\;\;\;\;\;\;x \in \Z;\;\;E>0 }
and subsequently for the $\R$ case:
\eqn\tarr{{1 \over \pi} \int_{-\infty}^{\infty} dq\
 e^{i q x} {1 \over E^2 + q^2}\
{E \over \sinh \pi E} = \  {1 \over \sinh \pi E}\  e^{-E |x|}
\;\;\;\;\;\;x \in \R;\;\;E>0 }
Quod erat demonstrandum. A more elegant way of showing
this result consists in identifying all momenta congruent zero
modulo two; thus ``periodizing'' the $\R$ propagator results
in the $\Z$ propagator:
\eqn\id{\sum_{n=-\infty}^{\infty} {E \over \sinh \pi E}
\ {1 \over E^2 + (p+2 n)^2}=
{\pi \over 2}\  {1 \over \cosh \pi E - \cos \pi p} }
It is interesting to compare the different pole structure in the
two propagators \ddiaa, \macroz. Each pole at some $E=i\nu$
corresponds to a on-shell (microscopic) state with wavefunction
$\mu^{{|\nu| \over 2}} K_{\nu}(\sqrt{\mu} l)$ satisfying
the Wheeler-DeWitt constraint. In \macror, \ddiaa the pole at $i\nu=iq$
signals the ``tachyon'' while the poles at $i\nu=ir$ are believed
to be related to redundant operators for generic $q$ and to
the special states for integer $q$ \msei. It is quite
curious that these poles {\it disappear } upon periodization and
are thus no longer present in the $\Z$ string, according to
 \id. In its place appear an infinite number
of gravitational descendents of the same structure
as those of the $C<1$ string models.

We will end this section by recalling how to extract
$n$--point functions from the correlation functions
of $n$ macroscopic loops \moore. One simply
shrinks the macroscopic loops and extracts the leading
non-analytic piece in the $\l_i$'s. Each macroscopic loop
turns into a local operator ${\cal O}(\l_i,p_i)$ regularized
by the loop-length $\l_i$.
In the case of the propagator
\macroz of the $\Z$-string this leads to the two--point function
\eqn\tpt{\langle {\cal O}(\l_1,p_1) {\cal O}(\l_2,p_2) \rangle =
-\delta^{(2)}(p_1 + p_2)
{1 \over |p|}[\Gamma(1-|p|)]^2 \mu^{|p|} \l_1^{|p|}
\l_2^{|p|} }
where $|p|<1$. After changing $p \rightarrow q$ and
$\delta^{(2)} \rightarrow \delta$ \tpt\ turns into
the same expression one obtains from the $\R$-string
propagator \macror, valid for $q \in \R$. This is of course
not accidental and will be discussed at the end of section 3.5..

\subsec{Three loops}

Considering the three-vertex we learn something about the interactions
in the two-dimensional string-theory. Here we have even less hope
to be able to perform the continuum path-integral (the moduli space becomes
quite complicated) than in the previous cases. Let us turn to our
two formulations of the string theory and see what we can understand
by comparing them. For the $\R$ string the diagram with three
external loops was calculated in \msei:
\eqn\thlpr{\eqalign{
\langle w_{q_1}(\l_1) w_{q_2}(\l_2) w_{q_3}(\l_3) \rangle =&
\delta(q_1 +q_2+q_3) {1 \over \mu}
 [\prod_{j=1}^{3} \int_{-\infty}^{\infty}dE_j
{E_j \over E_j -i q_{j}} K_{iE_j}(\sqrt{\mu} \l_i) ] \cr
& (E_1 + E_2 +E_3) \coth[{\pi \over 2}(E_1 + E_2 + E_3)] }}
For the $\Z$ string  the three-loop  correlator in
$p$-space (as discussed in the previous section) has been found
in \ivanlat  \ using the string theory Feynman rules:
\eqn\thlpz{\eqalign{
&\langle w_{p_1}(\l_1) w_{p_2}(\l_2) w_{p_3}(\l_3) \rangle =\cr
& =\delta^{(2)}(p_1 +p_2+p_3)
{1 \over \mu} \prod_{j=1}^{3} \int_{-\infty}^{\infty}dE_{j}
{E_{j} \over \sinh[{\pi \over 2} (E_{j}-ip_{j})]}
K_{iE_j}(\sqrt{\mu} \l_j) \cr}}
The above expression slightly differs from the one obtained
in \ivanlat . It is easy to see that if
the propagator in the Feynman rules of \ivanlat\ (eqs. (4.51), (4.53))
is replaced by its  by its chiral part
\eqn\chpr{{ E \ \sinh (\pi E) \cos (\pi p) \over
\cosh (\pi E) [\cosh (\pi E)- \cos (\pi p)]} \ \to \
{E \over \sinh (E-ip)}}
and the integration over $E$ is extended  over the whole real $E$-axis,
 the final result does not change.
Here and below we used the fact that the propagator is determined
completely by its poles and residues and one has the freedom to neglect
a factor which takes value  1 at all  poles.

Now we may argue as in the last section that the three loop correlators
of the two string models should coincide in target space.
It is not difficult to transform \thlpr,\thlpz\ back to
$x$-space; the resulting expressions are indeed identical.
Again the simplest way to demonstrate this is to periodize the function
\thlpr\  with respect  to the three momenta:
$q_{j} = p_{j}+ 2n_{j}; \ \ -1 \le p_{j} \leq 1, \ n_{j} \in \Z; \
j=1,2,3$.
 After that we introduce a Lagrange multiplier
to write the momentum-conservation $\delta$-function as
\eqn\lmmlm{
\delta \big[ \sum_{j=1}^{3}(p_{j}+2n_{j}) \big] =
\sum_{n\in \Z} \int _{0}^{2\pi} {d\alpha \over 2\pi}
 e^{i\alpha (n-n_{1}-n_{2}-n_{3})}
\delta [p_{1}+p_{2}+p_{3}-2n ]}
and apply the formula
\eqn\ffrm{\sum_{n\in \Z} {e^{-i\alpha n} \over E-i(p+2n)}
= {\pi \over 2} \  {e^{(\pi -\alpha ) (E-ip)/2} \over \sinh [\pi (E-ip)/2]}
, \ \ 0<\alpha <2\pi }
obtaining the r.h.s.of \thlpz\ up to a factor $\cos{\pi \over 2}
(p_{1}+p_{2}+p_{3})
\cosh{\pi \over 2}(E_{1}+E_{2}+E_{3})$ in the integrand which is equal to one
at all poles, q.e.d.

Let us discuss the pole structure of the two correlators \thlpr,\thlpz.
We see in the case of the $\R$ string, aside from the
tachyon poles at $i q_j$, infinitely many further, momentum
independent poles in \thlpr. They were attributed in \msei\ to
various contact terms of the macroscopic loops.
Once we restrict our target space to $\Z$
these contact terms {\it disappear} and we obtain \thlpz.
A rather dramatic consequence is that the removal of these
contact terms leads to a {\it factorization} of the interaction
which may be traced back to the fact that this interaction
takes place at a single point in $x$-space.

We conclude the section by comparing the three-point functions
obtained by shrinking the lenghts of the macroscopic loops
to zero. One easily finds from \thlpz\ for the $\Z$-string
\eqn\thpt{\langle \prod_{i=1}^{3} {\cal O}(\l_i,p_i) \rangle =
\delta^{(2)}(p_1+p_2+p_3) {1 \over \mu}
\prod_{i=1}^{3} \Gamma(1-|p_i|)({1 \over 2} \sqrt{\mu} \l_{i})^{|p_i|} }
Just as for the case of the two--point function this is,
upon replacing $p \rightarrow q$ and $\delta^{(2)} \rightarrow \delta$
the same analytical expression one obtains for the
$\R$-string (e.g. from eq. \thlpr).

\subsec{Four loops}

For the $\R$-string the correlator of four macroscopic loops
has not been calculated to our knowledge; in the SOS case the
result is quickly derived using the Feynman rules of the
string field theory \ivanlat.
There is a reducible $s$,$u$ and $t$ channel
diagram and an irreducible diagram. The $s$ channel diagram
reads
\eqn\sdia{\eqalign{
\delta^{(2)}(p_1+p_2 &+p_3+p_4)  ({1 \over 4 \mu})^2
[\prod_{j=1}^{4} \int_{-\infty}^{\infty} dE_j {1 \over \pi}
{E_j \over \sinh [\pi (E_j -i  p_j)/2]}
K_{i E_j}(\sqrt \mu \l_j)] \cr
&  \times \int_{-\infty}^{\infty} dE'
{E' \over \sinh [{\pi \over 2} ( E' - i (p_1 + p_2))]}
 } }
while the $t$ and $u$ channel diagrams are obtained by replacing
$ (p_1+p_2)$ by $ (p_1+p_3)$ and $ (p_1+p_4)$,
respectively. The irreducible diagram reads
\eqn\irdia{\eqalign{\delta^{(2)}(p_1+p_2+p_3+p_4)
{4 \over 9} ({1 \over 4 \mu})^2
[\prod_{i=j}^{4} \int_{-\infty}^{\infty} dE_j & {1 \over \pi}
{E_j \over \sinh [\pi (E_j - ip_j)/2]}
K_{i E_j}(\sqrt \mu \l_j)] \cr
&\times ({7 \over 4} + E_{1}^{2}+E_{2}^{2}+E_{3}^{2}+E_{4}^{2}) } }
After working out the integral in \sdia, shrinking the loops and combining
the four diagrams one obtains
\eqn\frpt{\eqalign{\langle \prod_{i=1}^{4} {\cal O}(\l_i,p_i) \rangle =
&\delta^{(2)}(p_1+p_2+p_3+p_4) {1 \over \mu^2}
\prod_{i=1}^{4} [\Gamma(1-|p_i|)({1 \over 2} \sqrt{\mu} \l_i)^{|p_i|}] \cr
&\times (|p_1+p_2| + |p_1+p_3| + |p_1+p_4| - 2) }}
This is precisely (after passing from $p$ to $q$ space and exchanging
$\delta^{(2)}$ for $\delta$) the
scattering amplitude obtained from the $\R$-string, valid for
all $q_j \in \R$.
Note that the cuts in \frpt\ stem entirely from the reducible
diagrams\foot{This is very reminiscent of the diagram technique of
DiFrancesco and Kutasov \dfkt\ . Our diagram technique is not identical,
but apparently closely related.}.

An interesting issue is whether correlation functions of macroscopic
loops as well as tachyonic microscopic operators in the
$\R$ theory may always (i.e. for any number of such
insertions and for arbitrary genus) be reconstructed from the $\Z$ theory.
We have argued that this should be possible for any given case by going
through target space. A more subtle issue is whether the
correlators of microscopic operators can always be obtained
by the simple replacements $p_j \rightarrow q_j$ and
$\delta^{(2)} \rightarrow \delta$. In fact it is straightforward to
prove that this has to work in the case of an $n$-point function
at least as long as $|q_1|+ \ldots +|q_n| < 2$.
We claim (but have not proven) that the resulting expression
can always be unambiguously continued to the
whole  $q$ space.

\newsec{Discussion}

The main purpose of the  present work is to establish the connection
between two alternative approaches to one dimensional non-critical
string theory: matrix quantum mechanics (``$\R$ strings'')
and the loop gas method (``$\Z$ strings'').
We have argued that correlators of macroscopic loops situated
at fixed points in the one dimensional target space (``punctual
boundary conditions'') should be identical. Note that the
argument given should apply to any number of macroscopic
loops as well as arbitrary genus. We have presented some explicit
examples at genus zero. We stressed the fact that in the
case of the $\Z$ string the macroscopic loops are confined to
sit at integer points in the target space. The momentum
space used to describe the scattering of $\Z$ strings is
therefore compact. This restricted scattering leads to rather
dramatic effects: The ``contact terms'' in the interactions
of the $\R$ strings become inobservable and all amplitudes factorize
into a simple set of elementary propagators and infinitely
many vertices corresponding to interactions
{\it local} in the one-dimensional target.
The amplitudes of the unrestricted target space
may nevertheless be reconstructed.

It would be interesting to directly prove our assertion for
arbitrary $n$-loop correlators  and any genus. This might
be difficult to do using the diagram technique; a possible
way to proceed might make use of a matrix model formulation
of the loop gas models which has recently been constructed \ivmm.

\bigbreak\bigskip\bigskip\centerline{{\bf Acknowledgements}}\nobreak
M.S. thanks SPhT Saclay for hospitality. This work was supported in
part by DOE grant DE-FG05-90ER40559.

\listrefs

\bye